\documentclass[10pt,pra,aps,showpacs,twocolumn,superscriptaddress]{revtex4-1}
\usepackage{eurosym}
\usepackage{lipsum}
\usepackage{graphicx}
\usepackage{amssymb}
\usepackage{amsmath}
\usepackage[usenames, dvipsnames]{color}
\usepackage[normalem]{ulem}
\usepackage{lipsum}
\usepackage{tikz} 

\begin{document}
\title{Vortex lattices in binary Bose-Einstein condensates with dipole-dipole interactions}
\author{Ramavarmaraja Kishor Kumar}
\affiliation{Instituto de F\'{i}sica, Universidade de S\~{a}o Paulo, 05508-090 S\~{a}o Paulo, Brazil}
\author{Lauro Tomio}
\affiliation{Instituto de F\'isica Te\'orica, Universidade Estadual Paulista, 01156-970 S\~ao Paulo, SP, Brazil}
\affiliation{InstitutoTecnol\'ogico de Aeron\'autica, DCTA,12.228-900 S\~ao Jos\'e dos Campos, SP, Brazil}
\author{Boris A. Malomed}
\affiliation{Department of Physical Electronics, School of Electrical Engineering, Faculty of Engineering,
Tel Aviv University, Tel Aviv 69978, Israel}
\affiliation{ITMO University, St. Petersburg 197101, Russia}
\author{Arnaldo Gammal}
\affiliation{Instituto de F\'{i}sica, Universidade de S\~{a}o Paulo,
05508-090 S\~{a}o Paulo, Brazil}

\date{\today}
\begin{abstract}
We study the structure and stability of vortex lattices in two-component rotating Bose-Einstein condensates with intrinsic dipole-dipole 
interactions (DDIs) and contact interactions. To address experimentally accessible coupled systems, we consider $^{164}$Dy-$^{162}$Dy 
and $^{168}$Er-$^{164}$Dy mixtures, which feature different miscibilities.
The corresponding dipole moments are $\mu_{\mathrm{Dy}}=10\mu _{\mathrm{B}}$ and $\mu _{\mathrm{Er}}=
7\mu _{\mathrm{B}}$, where $\mu _{\mathrm{B}}$ is the Bohr magneton. For comparison we also discuss a case 
where one of the species is non dipolar.
Under a large aspect ratio of the trap, we consider mixtures in the pancake-shaped format, which are modeled by effective 
two-dimensional coupled Gross-Pitaevskii equations, with a fixed polarization of the magnetic dipoles.
Then, the miscibility and vortex-lattice structures are studied, by varying the coefficients of the contact
interactions (assuming the use of the Feshbach-resonance mechanism) and the rotation frequency.
We present phase diagrams for several types of lattices in the parameter plane of the rotation frequency and 
ratio of inter- and intra-species scattering lengths.  
The vortex structures are found to be diverse for the more miscible $^{164}$Dy-$^{162}$Dy mixture, with a variety
of shapes, whereas for the less miscible case of $^{168}$Er-$^{164}$Dy, the lattice patterns mainly feature circular or square 
formats. 
\end{abstract}
\maketitle
\section{Introduction}
\label{secI} 
In Bose-Einstein condensates (BECs), quantized vortices emerge above a certain critical rotation frequency~\cite{review},
which may be imposed by techniques such as rotating traps, laser stirring, and the addition of an oscillatory excitation to the 
trapping potential~\cite{Exp}. Experiments for vortices have also been performed with multicomponent BECs. In this regard, 
we can mention Ref.~\cite{coreless}, as well as works cited in the recent review by Martin et al.~\cite{Parker}.
In particular, the study on vortices in binary condensates are interesting due to the fact that inter-species interactions
produce diverse vortex structures in addition to the fundamental Abrikosov's triangular lattice, such that squared-shaped, 
coreless, with domain walls, droplets, as well as isolated density peaks in the two-component 
mixtures~\cite{Mueller,vortex-phase,vort-sheet,2c-vort}.

Recent experiments with ${}^{168}$Er and $^{164}$Dy condensates also
stimulate interest to properties of quantum droplets that can be created in
dipolar BECs~\cite{droplet1, droplet2, droplet-er}.
These studies have revealed that the droplets are self-trapped as many-body states
in bosonic gases, supported by the balance between attractive and repulsive
forces in these settings. With the attractive forces being provided by
dipole-dipole interactions (DDIs), the repulsive cases are induced by the
beyond-mean-field quantum fluctuation effects, known as Lee-Huang-Yang
corrections (LHY). LHY corrections are also used to 
predict stable droplets in non-dipolar two-component systems~\cite{Petrov1,Petrov2,2017-Li}, which have 
been created very recently in experiments~\cite{Leticia1,Inguscio,Leticia2}.

The objective of this paper is to study rotational regimes of two-component
dipolar BECs. Previous studies dealing with vortices in dipolar BECs have
explained the role of dipole-dipole interactions (DDI) in the formation of
vortices, as reported in the review \cite{Parker}. In particular, the theoretical
analysis dealt with the calculation of the critical rotation frequency and
vortex structures under the action of the DDI~\cite{Yi2006,Malet2011,Kishor-vort}. 
In the two-component setting with one dipolar component and the other one carrying no 
dipole moments, the dipolar component features a smaller critical rotation frequency than 
its non-dipolar counterpart, and the dipolar component produces a larger number of vortices 
than the non-dipolar one, at the same rotation frequency~\cite{2c-compet-dip}.

In general, vortex states in binary BECs are strongly affected by the (im)miscibility. 
Completely miscible settings feature triangular, square-shaped, and rectangular vortex 
lattices, depending on the rotation frequency. On the other hand, immiscible binary 
condensates support bound states of vortices, stripes, and vortex sheets (domain 
walls)~\cite{Mueller, vortex-phase}. In this regard, to explore vortex structures in binary 
condensates, we rely on results obtained in a previous study, reported in Ref.~\cite{2017jpco},
for the miscibility of two-component dipolar systems. In this reference, by considering stability
requirements and miscibility properties, it was found more appropriate to consider 
pancake-shaped symmetry for the trap, varying the inter- and intra-species contact 
interactions to modify the miscibility properties.

In the present work, we report manifestations of the miscibility-immiscibility transition in dipolar
mixtures in terms of vortex-lattice configurations. We address vortex-lattice states in two-component 
BECs under the action of the DDI, by considering the same dipolar systems previously studied in 
Ref.~\cite{2017jpco}. Following that, our analysis is performed for a pancake-shaped trap configuration, 
in which the underlying system of  three-dimensional (3D) coupled Gross-Pitaevskii equations (GPEs) 
can be reduced to a two-dimensional (2D) form.

First, we consider the system with parameters corresponding to a nearly
symmetric (with respect to the two components) $^{164}$Dy-$^{162}$Dy
mixture, where both components have equal dipole moments, thus supporting
the balance between intra- and inter-species DDI. In the absence
of contact interactions, this mixture is miscible. Its
miscibility-controlling parameter is the ratio between scattering lengths of
contact inter- and intra-species repulsion, if it is added to the DDI. Next, we consider 
the setting with parameters of the
asymmetric $^{168}$Er-$^{164}$Dy mixture, with unequal dipole moments in the
two species, which gives rise to imbalance of the inter- and intra-species
DDI. This dipolar mixture produces immiscible states, in the
absence of contact interactions. In this case too, mixing-demixing
transition is controlled by the ratio of the scattering lengths of intra-
and inter-species contact interactions, if they are present. The miscibility
of these binary dipolar condensates determines vortex-lattice structures
which can be created in them. We also consider briefly another binary
system, in which only one component carries dipole moment. The latter
system favors the miscibility, in comparison with the symmetric and
asymmetric mixtures of two dipolar components.

The paper is organized as follows. In Sec.~\ref{secII}, we present the 2D
mean-field model for the trapped two-component dipolar BEC under rotation,
and numerical methods used in this work. In Sec.~\ref{secIII}, we report
numerical results for phase diagram of vortex-lattice patterns, varying the
strength of inter- and intra-species contact interactions versus the
rotation frequency, for different dipolar mixtures. Some analytical results,
based on the Thomas-Fermi approximation for the immiscible system, are
presented too. The paper is concluded by Sec.~\ref{secIV}.

\section{Coupled dipolar condensates under rotation}
\label{secII} The system of coupled GPEs for the binary condensate with 
the DDI, for the two component wave-functions 
$\Psi _{j=1,2}\equiv \Psi _{j}({\mathbf{r}},t)$, can written as~\cite{goral2002,2012-Wilson} 
{\small
\begin{eqnarray}
\mathrm{i}\hbar \frac{\partial \Psi _{j}}{\partial t} &=&
\bigg [-\frac{\hbar ^{2}}{2m_{j}}\nabla ^{2}+V_{j}({\mathbf{r}})-\overline{\Omega}\hbar L_{z}+
\sum_{k=1}^{2}G_{jk}N_{k}|\Psi _{k}|^{2}  \notag
\\
&+&\sum_{k=1}^{2}\frac{N_{k}}{4\pi }\int d^{3}{\mathbf{r}^{\prime }}
V_{jk}^{(d)}({\mathbf{r}-\mathbf{r}^{\prime }})|\Psi _{k}^\prime|^{2}{\bigg ]}\Psi _{j},  \label{eq_DBEC1}
\end{eqnarray}} 
where $\Psi _{k}^\prime\equiv  \Psi _{k}({\mathbf{r^\prime}},t)$.
The masses, number of atoms and trapping potentials for the two species $j$ are, respectively
given by $m_{j}$, $N_{j}$ and $V_{j}(\mathbf{r})$. 
Further, $V_{jk}^{(d)}(\mathbf{r-r^{\prime }})$ are the kernels of the DDI, 
$\overline{\Omega}$ a common rotation frequency of both components, with
$\hbar L_{z}=-\mathrm{i}\hbar(x\partial /\partial {y}-y\partial /\partial {x})$ being
the angular-momentum operator. The strengths of the contact interactions are $G_{jk}\equiv ({2\pi \hbar
^{2}}/{m_{jk}})a_{jk}$, where $m_{jk}=m_{j}m_{k}/(m_{j}+m_{k})$ are the
respective reduced masses, while $a_{jk}$ are the corresponding intra- ($a_{jj}$) and inter-species 
($a_{12}$) two-body scattering lengths. Trapping
is provided by Harmonic-oscillator (HO) potentials with frequencies $\omega_{j}$,
\begin{equation}
V_{j}({\mathbf{r}})=\frac{1}{2}m_{j}\omega _{j}^{2}(x^{2}+y^{2}+\lambda
^{2}z^{2}),  \label{trap}
\end{equation} 
and common aspect ratio $\lambda $ for both components, such that the trap
is spherically symmetric for $\lambda =1$, cigar-shaped for $\lambda <1$,
and pancake-shaped for $\lambda >1$. The DD-interaction kernels in Eq. (\ref 
{eq_DBEC1}) correspond to the configuration with dipole moments polarized
(by an external magnetic field) perpendicularly to the $\left( x,y\right) $
plane:
\begin{equation}
V_{ij}^{(d)}(\mathbf{r-r^{\prime }})=D_{ij}\frac{1-3\cos ^{2}\theta }{| 
\mathbf{r-r^{\prime }}|^{3}},  \label{DDI}
\end{equation} 
where $\theta $ is the angle between the polarized magnetic moments and $ 
\left( \mathbf{r-r^{\prime }}\right) $, and $D_{ij}\equiv \mu_{0}\mu_{i}\mu _{j}$, 
with the free-space permeability $\mu _{0}$. 

For the pancake-shaped dipolar BEC $(\lambda\gg 1)$, we assume the
usual factorization of the wave function into the ground state of the
transverse HO trap and a 2D wave function, as
\begin{equation}
\Psi _{j}(\mathbf{r},t)=\frac{1}{(\pi d_{z}^{2})^{1/4}}\exp \left( -\frac{
z^{2}}{2d_{z}^{2}}\right) \Phi _{j}(x,y,t),  \label{ansatz1}
\end{equation}
where $d_{z}\equiv \sqrt{1/\lambda }$ is the trap's HO length~\cite{Luca,Luca2,CPC2,2012-Wilson}.
To derive the effective coupled GPEs in 2D, we insert the ansatz (\ref{ansatz1}) in Eq. (\ref
{eq_DBEC1}), multiplying the equation by another power of the HO
ground-state wave function, performing integration over $z$. The coupled
equations are cast in a dimensionless format by measuring the energy and
length in units of $\hbar \omega _{1}$ and $l\equiv \sqrt{\hbar
/(m_{1}\omega _{1})}$, respectively. 
By taking $x, y$ variables in units of $l$ ($x\to l x$ and $y\to l y$), the accordingly rescaled 
quantities are
\begin{eqnarray}
{\boldsymbol{\rho}}&\equiv&{x}\hat{e}_1+{y}\hat{e}_2;\;\;
 \tau\equiv\omega_1 t,\nonumber\\
g_{jk}&\equiv& \sqrt{2\pi \lambda }\frac{m_{1}}{m_{jk}}\frac{a_{jk}N_{k}}{l}
,\;\;\;\sigma \equiv \frac{m_{2}\omega _{2}^{2}}{m_{1}\omega _{1}^{2}},
\nonumber\\
a_{jj}^{(d)}&\equiv& \frac{D_{jj}}{12\pi }\frac{m_{j}}{m_{1}}\frac{1}{
\hbar \omega _{1}l^{2}},\;\;\;a_{12}^{(d)}=a_{21}^{(d)}=\frac{D_{12}}{12\pi }
\frac{1}{\hbar \omega _{1}l^{2}},\label{par}\\
\mathrm{d}_{jk}&=&\frac{N_{j}D_{jk}}{4\pi }\frac{1}{\hbar \omega _{1}\,l^{3}}
\;\;\; (j,k = 1,2)
\nonumber,
\end{eqnarray}
with the corresponding 2D wave function for the components $j=1,2$ given by
\begin{equation}
\psi_j(\boldsymbol{\rho},\tau) \equiv l\, \Phi_j(x,y,t).
\end{equation}

In terms of this notation, for 
$\psi_{j}\equiv \psi_{j}(\boldsymbol{\rho },\tau)$ and
$\psi_{j}^\prime\equiv \psi_{j}(\boldsymbol{\rho}^\prime,\tau)$
the coupled GPEs in 2D take the form of
{\small \begin{eqnarray}
\mathrm{i}\frac{\partial \psi _{1}}{\partial \tau }&=&
\bigg[-\frac{1}{2}{\nabla _{\boldsymbol{\rho }}^{2}}+\frac{\rho^{2}}{2}-\Omega L_{z}+g_{11}|\psi _{1}  |^{2}+
g_{12}|\psi _{2}  |^{2}  \nonumber \\
&+&\int d\boldsymbol{\rho }^{\prime }V^{(d)}(\boldsymbol{\rho }-\boldsymbol{\rho }^{\prime })
\left( \mathrm{d}_{11}|\psi _{1}^\prime   |^{2}+\mathrm{d}_{12}
|\psi _{2}^\prime   |^{2}\right)\bigg]\psi _{1}, \notag\\ \label{2d-2c} \\
\mathrm{i}\frac{\partial \psi _{2}  }{\partial \tau }
&=&\,{\bigg [}-\frac{m_{1}}{2m_{2}}{\nabla _{\boldsymbol{\rho}}^{2}}+
\frac{\sigma\,\rho^{2}}{2}-\Omega L_{z}+g_{22}|\psi _{2}  |^{2}+g_{21}
|\psi _{1}  |^{2} \notag \\
&+&\int d\boldsymbol{\rho }^{\prime }V^{(d)}(\boldsymbol{\rho }-\boldsymbol{\rho }^{\prime })
\left( \mathrm{d}_{22}|\psi _{2}^\prime   |^{2}+\mathrm{d}_{21}
|\psi _{1}^\prime   |^{2}\right) {\bigg ]}\psi _{2}  ,  \notag
\end{eqnarray}}
where the common rotation frequency of the two components was written in terms 
of $\omega_1$, such that $\Omega=\overline{\Omega}/\omega_1$.

In Eq. (\ref{2d-2c}) the DD-interaction terms can be expressed by
means of the convolution theorem,
{\small \begin{equation}
\sum_{j=1}^{2}\int d\boldsymbol{\rho }^{\prime }V^{(d)}(\boldsymbol{\rho }-
\boldsymbol{\rho }^{\prime })|\psi _{j}^{\prime}|^{2}
=\mathcal{F}_{2D}^{-1}\left[ \tilde{V}^{(d)}(k{_{\rho }})\tilde{n}_{j}({\mathbf{k}_{\rho }},\tau)\right] ,
\end{equation}}
where $\mathcal{F}_{2D}^{-1}$ is the inverse 2D Fourier-transform operator, 
with  $k_{\rho }\equiv \sqrt{k_{x}^{2}+k_{y}^{2}}$, 
{\small
\begin{equation}
\widetilde{n}_{j}(\mathbf{k}_{\rho },\tau)=\int d\boldsymbol{\rho }e^{i\mathbf{k
}_{\rho }.\boldsymbol{\rho }}|\psi _{j}  |^{2},~
\widetilde{n}_{j}(k_{z})=e^{-k_{z}^{2}d_{z}^{2}/4},
\end{equation}}
and 
{\small \begin{gather}
\tilde{V}^{(d)}(k{_{\rho }})\equiv \frac{1}{2\pi }\int_{-\infty }^{\infty
}dk_{z}\left( \frac{3k_{z}^{2}}{\mathbf{k}^{2}}-1\right) |\widetilde{n}
_{j}(k_{z})|^{2}  \label{V} \\
\,=\frac{1}{\sqrt{2\pi }d_{z}}\left[ 2-\frac{3\sqrt{\pi }}{\sqrt{2}}k_{\rho
}d_{z}\exp \left( \frac{k_{\rho }^{2}d_{z}^{2}}{2}\right) \left\{ 1-\mathrm{
erf}\left( \frac{k_{\rho }d_{z}}{\sqrt{2}}\right) \right\} \right] .  \notag
\end{gather}}

For the numerical solution of Eq.~(\ref{2d-2c}), we employed the split-step
Crank-Nicolson method, as in Refs.~\cite{Gammal2006,CPC1,CPC2}, combined
with the standard method for evaluating DD-interaction integrals in the
momentum space~\cite{CPC2,goral2002}. To look for stable solutions,
numerical simulations were carried out in imaginary time on a grid with $512$
points in $x$ and $y$ directions, spatial steps $\Delta x=\Delta y=0.05$ and
time step $\Delta t=0.0005$. Both component wave functions are 
renormalized to one, $\int d\boldsymbol{\rho }|\psi _{j}|^{2}=1$, at each time step.

To calculate stationary vortex states, Eq.~(\ref{2d-2c}) was solved with
different initial conditions. From the tests, we choose the following
suitable initial conditions in the form of a combination of angular
harmonics~\cite{Rokhsar},
\begin{equation}
\psi_j (\boldsymbol{\rho },0)=\sum_{m=0}^{L}\frac{( x+{\mathrm{i}}y)^{m}
\, e^{(-{\rho^{2}}/{2})}}
{\sqrt{\pi (L+1)m!}}
\exp\left(2\pi {\mathrm{i}}\mathcal{R}_m\right),  \label{initial}
\end{equation}
\noindent where $\mathcal{R}_m$ is a randomly generated number distributed uniformly between $0$ 
and $1$, with arbitrary integer value for $L$ that we have consider up to $L=40$. 
In addition, we checked the convergence of the solutions with inputs as considered 
in Ref.~\cite{Bao}.

For the parameters, we follow the ones used in a previous study on miscibility in 
coupled dipolar condensates, given in Ref.~\cite{2017jpco}, for these atomic mixtures with 
Erbium ($^{168}$Er) and Dysprosium ($^{162,164}$Dy) isotopes. In terms of the Bohr 
magneton $\mu_B$, the corresponding assumed dipole moments are, respectively,  
$\mu=7\mu_B$ and $\mu=10\mu_B$. 
In the harmonic axial traps, defined in Eq.~(\ref{trap}), the assumed angular frequencies 
confining each species were such that 
$\omega_j = 2\pi\times 60 {\rm s}^{-1}$ for the $^{168}$Er and 
$\omega_j = 2\pi\times 61 {\rm s}^{-1}$ for the $^{162,164}$Dy,
such that $\sigma$ defined in Eq.~(\ref{par}) is close to one.
The time and space units are such that $1/\omega_1=$ 2.65 ms and $l = 1\mu$ m 
($= 10^4\AA$ = 1.89$\times 10^4 a_0$).
As found appropriate for experimentally realistic settings, in all the following analysis and
results we are taking a pancake-shaped 
trap, with an aspect ratio $\lambda =20$, and fix the number of atoms to be equal for both
species with $N_{1}=N_{2}=10^{4}$. 
The contact and dipole-dipole interactions, expressed in terms of the Bohr radius $a_0$,
are being varied by considering several conditions of interest in view of miscibility
properties of the binary mixtures. In particular, as considering the corresponding 
dipole moments, 
the strengths of the DD interaction are given as $a_{11}^{(d)}=a_{22}^{(d)}=131\,a_{0}$ 
and $a_{12}^{(d)}=a_{21}^{(d)}=131\,a_{0}$, for the $^{164}$Dy-$^{162}$Dy mixture;  and 
$a_{11}^{(d)}=66\,a_{0} $, $a_{22}^{(d)}=131\,a_{0}$ and $a_{12}^{(d)}=a_{21}^{(d)}=94\,a_{0} $, 
for the $^{168}$Er-$^{164}$Dy mixture.
Further, in order to explore various families of vortex patterns, we varied the rotation 
frequency $\Omega $.

As transitions between vortex-lattice structures are determined by the
miscibility, it is appropriate to consider a parameter to measure the 
overlapping between densities of the components, as the one defined in 
Ref.~\cite{2017jpco}:
\begin{equation}
\eta =\int |\psi _{1}| |\psi _{2}| d\boldsymbol{\rho }\equiv 
\int \sqrt{|\psi _{1}|^{2}|\psi_{2}|^{2}}\ d\boldsymbol{\rho }.  \label{eta}
\end{equation}
As $\psi _{1}$ and $\psi _{2}$ are both normalized to $1$, a complete
overlap between the species have $\eta =1$, with indications of partial
overlapping for smaller values of $\eta$. 
Results reported in the following are suggesting that values of $\eta \lesssim 0.5$
correspond to almost clear demixing, as the maxima for the densities are located at
well-separated points. In the interval of $0.5<\eta <0.8$, the system may be
 categorized as a partially miscible, as one can notice that the peaks of the densities
are approaching each other. The density maxima are nearly overlapping for 
$\eta \gtrsim 0.8$, such that we can identify the system as a miscible one.
The applicability of these definitions was checked for all settings considered in this work.

\section{Results}
\label{secIII}
The numerical results presented in this section are organized in four subsections, considering
possible nonlinear effects due to the interplay of the contact and dipolar interactions.
We start by considering a pure-dipolar case (subsection A). Next, in other subsections, we 
vary the strength of contact interactions, considering a nearly symmetric dysprosium-dysprosium 
mixture (subsection B); a non-symmetric erbium-dysprosium mixture (subsection C); and, finally, a
mixture of dipolar and non-dipolar species (subsection D). As mentioned above, the harmonic-trap 
aspect ratio and number of atoms in both components are fixed, respectively, to $\lambda=20$ and 
$N_1=N_2=10^4$, which are adjusted to the previous studies of stability and miscibility of binary 
dipolar condensates~\cite{2017jpco}. All the following results are 
produced with parameters of the the contact and dipolar interactions given in units of the Bohr radius 
$a_0$. Adopting the length unit as $l=1.89\times 10^4 a_0$, the
coordinates and densities are presented as dimensionless quantities. 

\subsection{Vortex structures in dipolar binary condensates in the absence of contact interactions}
\begin{figure}[tbp]
\includegraphics[width=0.5\textwidth]{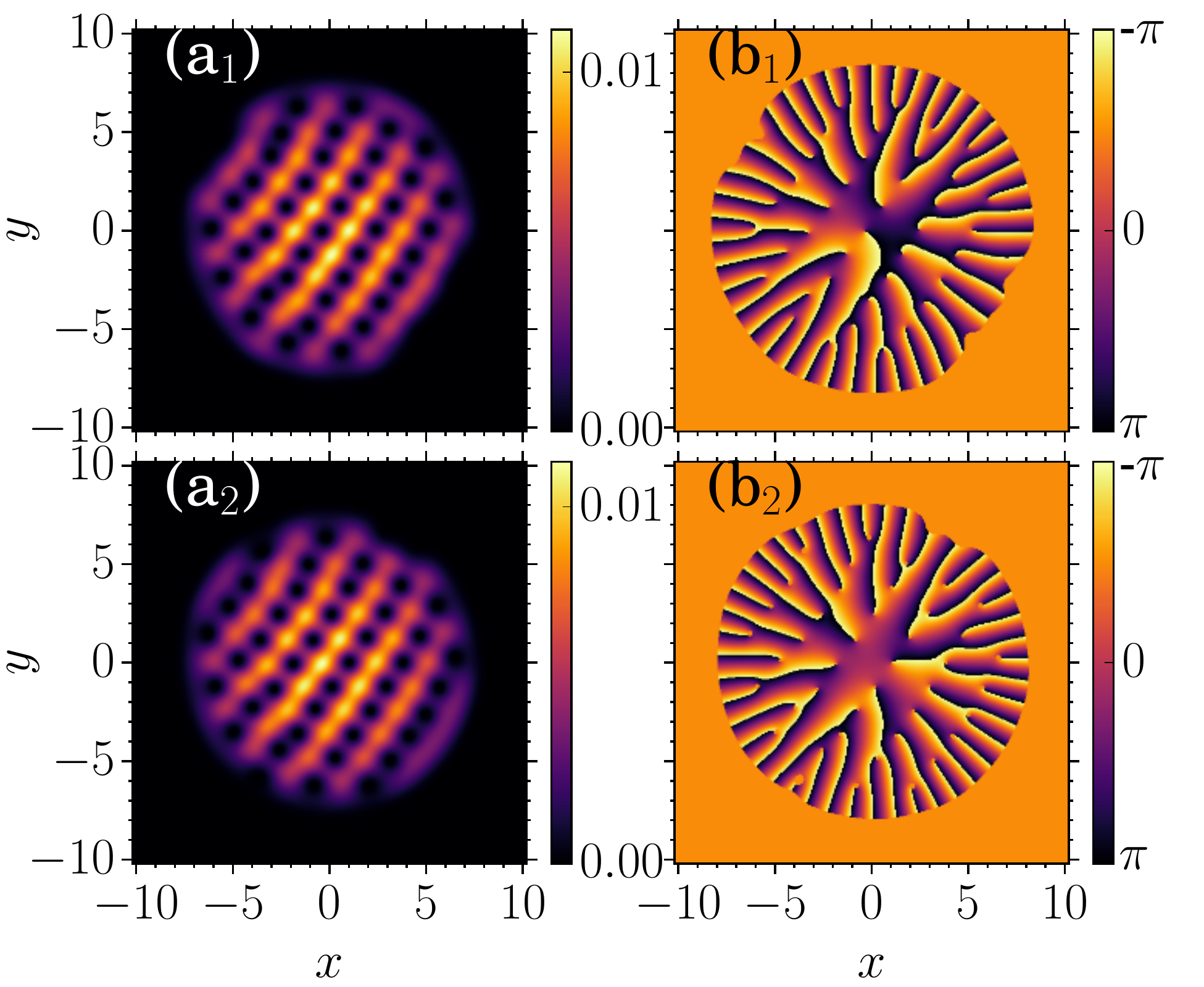}
\caption{The 2D density (left frames) and phase (right frames) patterns for the miscible $^{164}$Dy-$^{162}$Dy system
with no-contact interactions are shown by the upper (component $j=1$) and lower (component $j=2$) panels. 
The parameters are: $a_{jk}=0$, $a_{jk}^{(d)}=131a_{0}$, $\Omega =0.6$, $\protect\lambda =20$, and $N_{j}=10^{4}$
($j,k = 1,2$).
}
\label{fig1}
\end{figure}

\begin{figure}[tbp]
\includegraphics[width=0.5\textwidth]{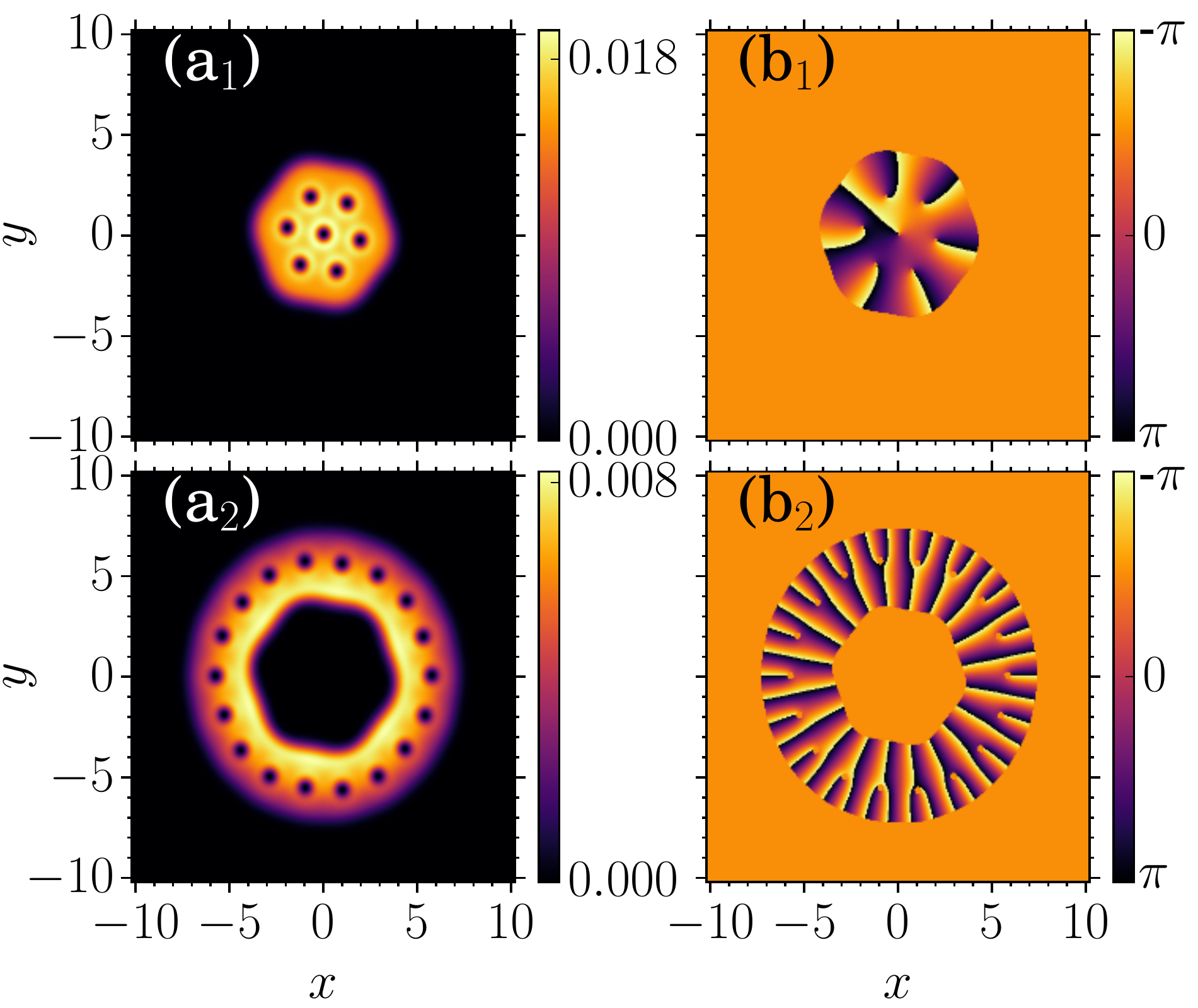}
\caption{The 2D density (left frames) and phase (right frames) patterns, 
for the immiscible $^{168}$Er-$^{164}$Dy system, with no-contact interactions are
shown by the upper (component $j=1$) and lower (component $j=2$) frames.
The parameters are: $a_{jk}=0$, $a_{11}^{(d)}=66a_{0} $, $a_{22}^{(d)}=131a_{0}$, 
$a_{12}^{(d)}=a_{21}^{(d)}=94a_{0}$, $\Omega =0.6$, $\protect\lambda =20$ and 
$N_{j}=10^{4}$ ($j,k = 1,2$).}
\label{fig2}
\end{figure}

To illustrate the miscible/immiscible states in the absence of the contact interactions 
($a_{jk}=0$, for $j, k = 1,2$), we display stable solutions for densities and phases, 
corresponding to the dipolar mixtures $^{164}$Dy-$^{162}$Dy (Fig.~\ref{fig1}) 
and $^{168}$Er-$^{164}$Dy (Fig.~\ref{fig2}). In both the cases, we apply the same 
rotation frequencies $\Omega =0.6$ and aspect ratios $\lambda=20$.
As defined by Eq.~(\ref{eta}), the miscibility parameter is much larger for the 
$^{164}$Dy-$^{162}$Dy BEC mixture, $\eta =0.81$, corresponding to almost 
completely miscible state. On the other hand, for the $^{168}$Er-$^{164}$Dy system, 
we have a smaller value of $\eta =0.19$, implying in an almost immiscible mixture. 
The predicted lattice patterns for the vortices, considering these miscible and immiscible 
binary dipolar condensates, are presented in Figs.~\ref{fig1} and \ref{fig2}, respectively. 
The patterns may be naturally classified as from squared- to striped-shaped lattices in 
the more miscible case, whereas as having finite segment of a hexagonal lattice in one 
component, surrounded by a ring in the other component in the immiscible mixture.

\begin{figure}[h]
\includegraphics[width=0.45\textwidth]{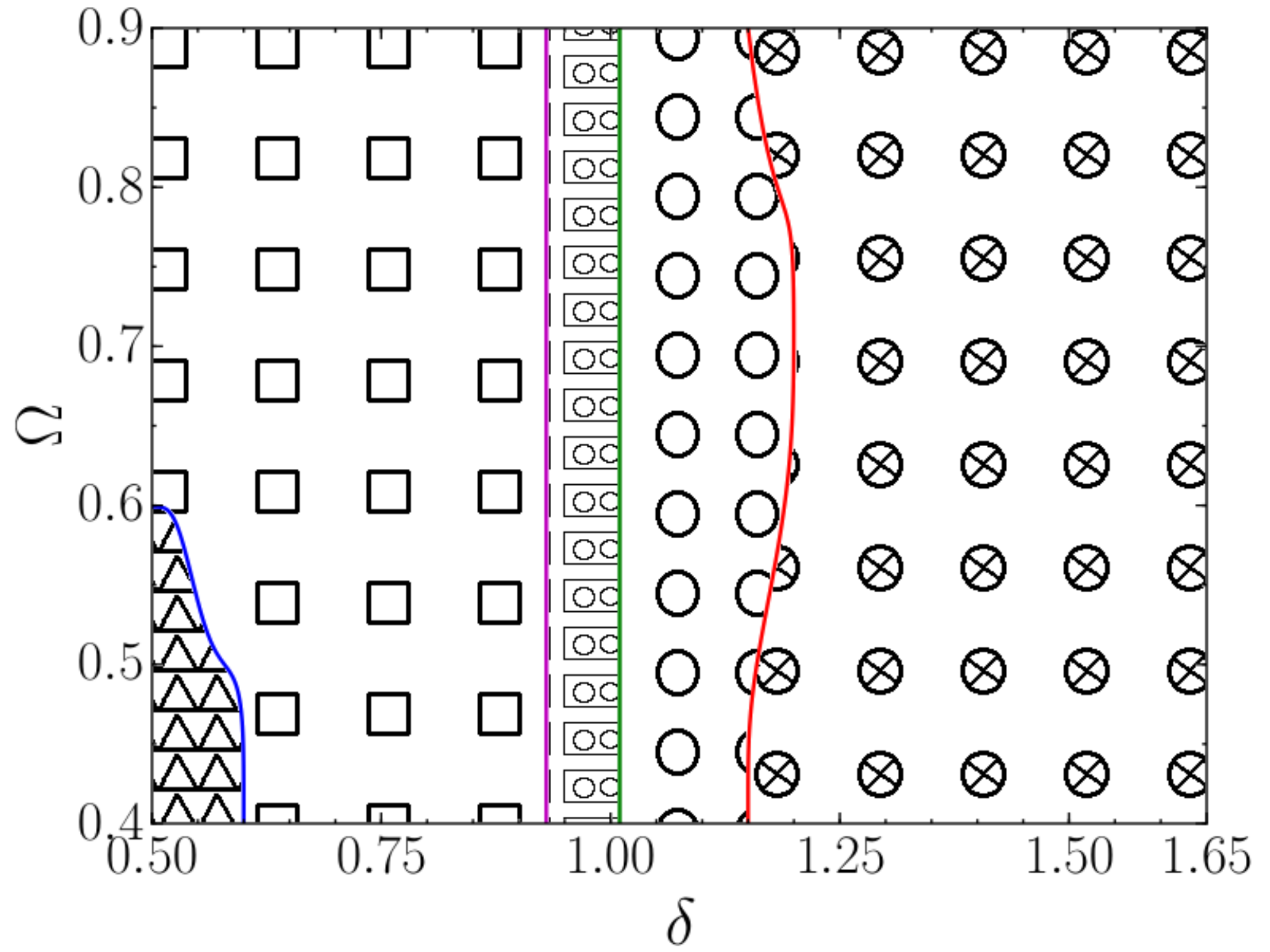}
\caption{The phase diagram of vortex patterns in the plane defined by the angular
velocity $\Omega$ and the ratio between inter- and intra-species scattering length
$ \protect \delta\equiv a_{12}/a_{11}$, for the $^{164}$Dy-$^{162}$Dy mixture.
The symbols for the observed patterns are: triangles for triangular lattices; squares for 
squared lattices; double circles inside a rectangle for rectangular or double-core vortices; 
circles for striped vortices; crossed circles for domain walls. Typical examples are 
displayed in Figs.~\protect\ref{fig4}-\protect\ref{fig6}.
}
\label{fig3}
\end{figure}

\subsection{The nearly symmetric $^{164}$Dy-$^{162}$Dy mixture under the
action of contact interactions}

To drive the mixing-demixing transition in the dipolar mixtures in the presence of contact 
interactions, we use the scattering lengths as tuning parameters. For that, the intra-species 
scattering lengths are assumed to be equal, $a_{11}=a_{22}$, with the ratio between inter- 
and intra-species scattering lengths being defined by
\begin{equation}
\delta =\frac{a_{12}}{a_{11}}.
\label{delta}
\end{equation}
Therefore, stable vortex states are explored by varying this ratio $\delta$ and the
rotation frequency $\Omega $, which is given in units of the trap frequency
$\omega_1$. This parameter for the rotation we consider in the interval $0.4<\Omega<0.9 $, 
as this interval adequately represents various types of vortex patterns which the system can 
generate.
In this work we restrict the analysis to the case of equal intra-species scattering lengths, 
$a_{11}=a_{22}$, as effects produced by the variation of the inter/intra-interaction ratio 
$\delta$, defined by Eq. (\ref{delta}), are essentially stronger (and more interesting) than 
what may be controlled by the variation of $a_{22}/a_{11}$. Systematic analysis of the
latter effects may be considered separately, to keep the length of the present paper in 
reasonable limits.

\begin{figure}[h] 
\includegraphics[width=0.45\textwidth]{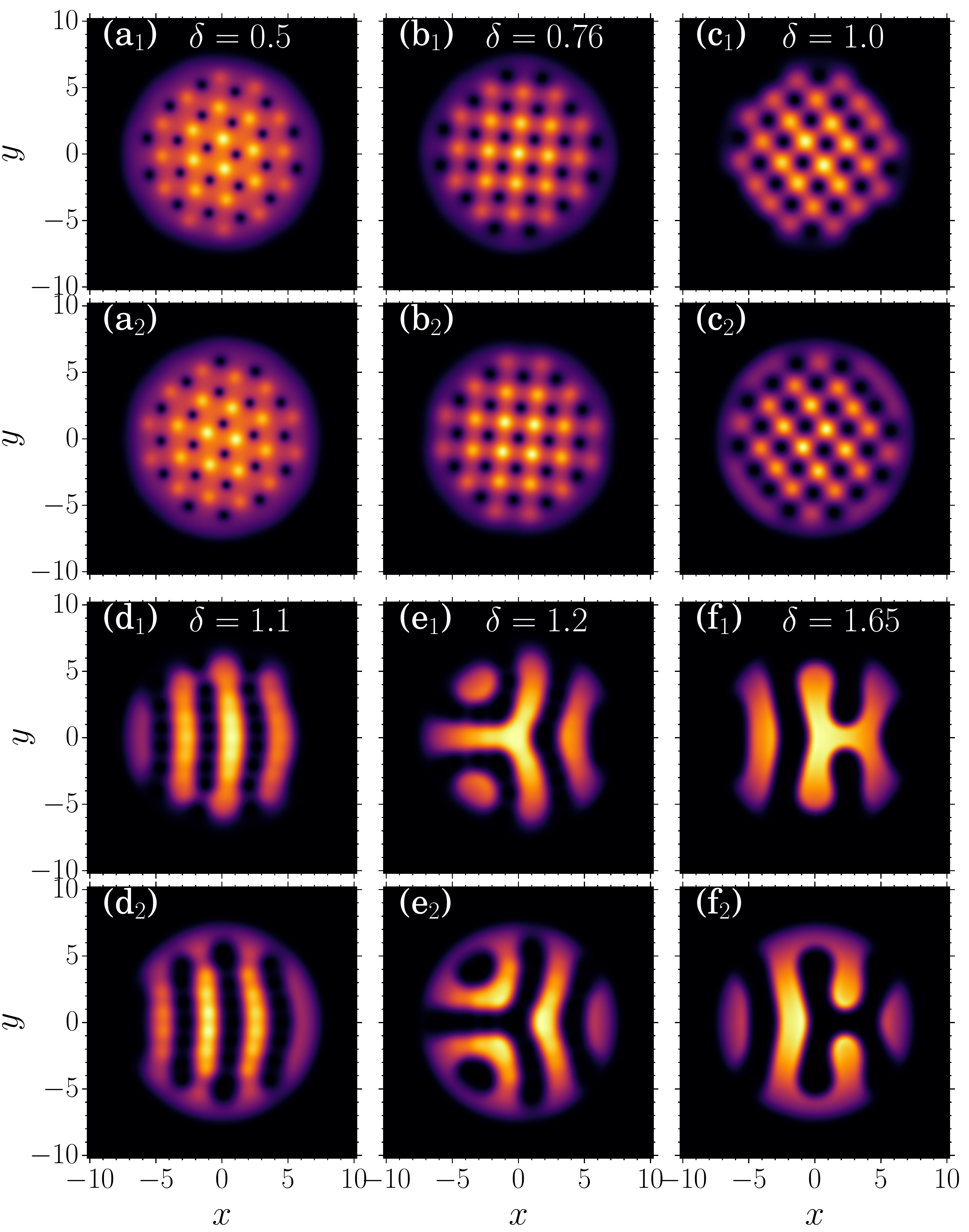}
\caption{2D component densities, $|\protect\psi_{j}|^{2}$ [(a$_j$) to (f$_j$), with $j=$1,2], 
for the $^{164}$Dy-$^{162}$Dy mixture in different patterns of stable vortices, following 
phase diagram Fig.~\ref{fig3} for $\protect\delta$ varying from 0.5 to 1.65, 
as indicated. The lattices are triangular (a$_{j=1,2}$), squared
(b$_{j=1,2}$), rectangular (c$_{j=1,2}$),  striped (d$_{j=1,2}$), and with domain walls
[(e$_{j=1,2}$) and (f$_{1,2}$)]. Other parameter are: $N_{j=1,2}=10^{4}$, $\lambda =20$, 
$a_{12}=50a_{0}$, and $\Omega =0.4$.}
\label{fig4}
\end{figure}

A commonly known result for spatially uniform states in the absence of DD
interactions is that the miscibility and immiscibility take place at $\delta
<1$ and $\delta >1$, respectively \cite{mineev}. Stable vortex structures
found in different domains of the $\left( \delta ,\Omega \right) $ plane for
parameters of the $^{164}$Dy-$^{162}$Dy mixture (assuming that $a_{12}$ can
be adjusted by means of the Feshbach resonance) are summarized in the phase
diagram exhibited in Fig.~\ref{fig3}, with typical examples of different
stable patterns shown in Fig.~\ref{fig4} for $\Omega =0.4$.

\begin{figure*}[tbp]
\centering \includegraphics[width=0.8\textwidth]{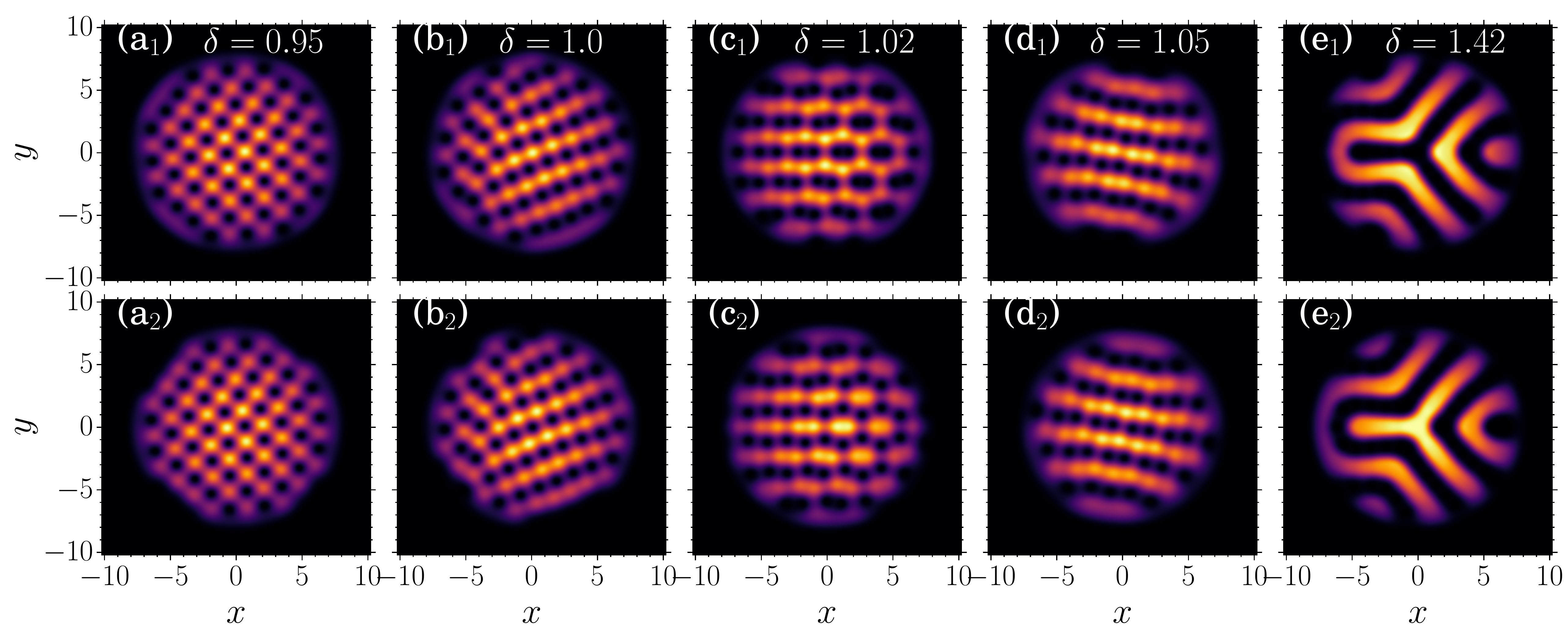}
\caption{The 2D component densities, with same parameters as in Fig.~\ref{fig4}, by changing the rotation 
frequency to $\Omega=0.6$.
}
\label{fig5}
\end{figure*}

\begin{figure}[tbp]
\centering\includegraphics[width=0.5\textwidth]{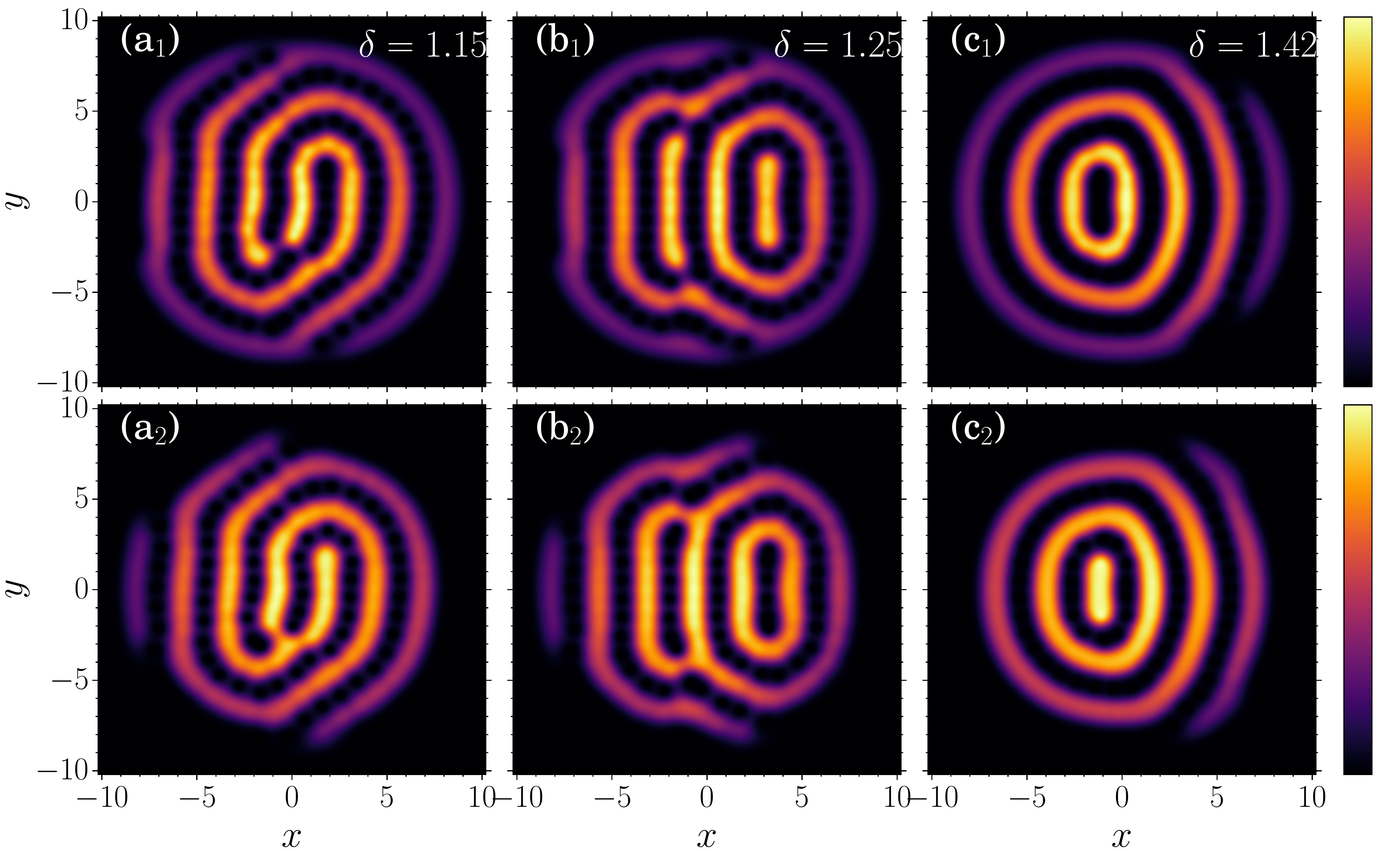}
\caption{The 2D component densities, with same parameters as in Fig.~\ref{fig4}, by changing the rotation 
frequency to $\Omega=0.8$. 
The color bar on the right specifies variation of the density from 0 (darker) to 0.01
(lighter).}
\label{fig6}
\end{figure}

\begin{figure}[tbp]
\centering\includegraphics[width=0.4\textwidth]{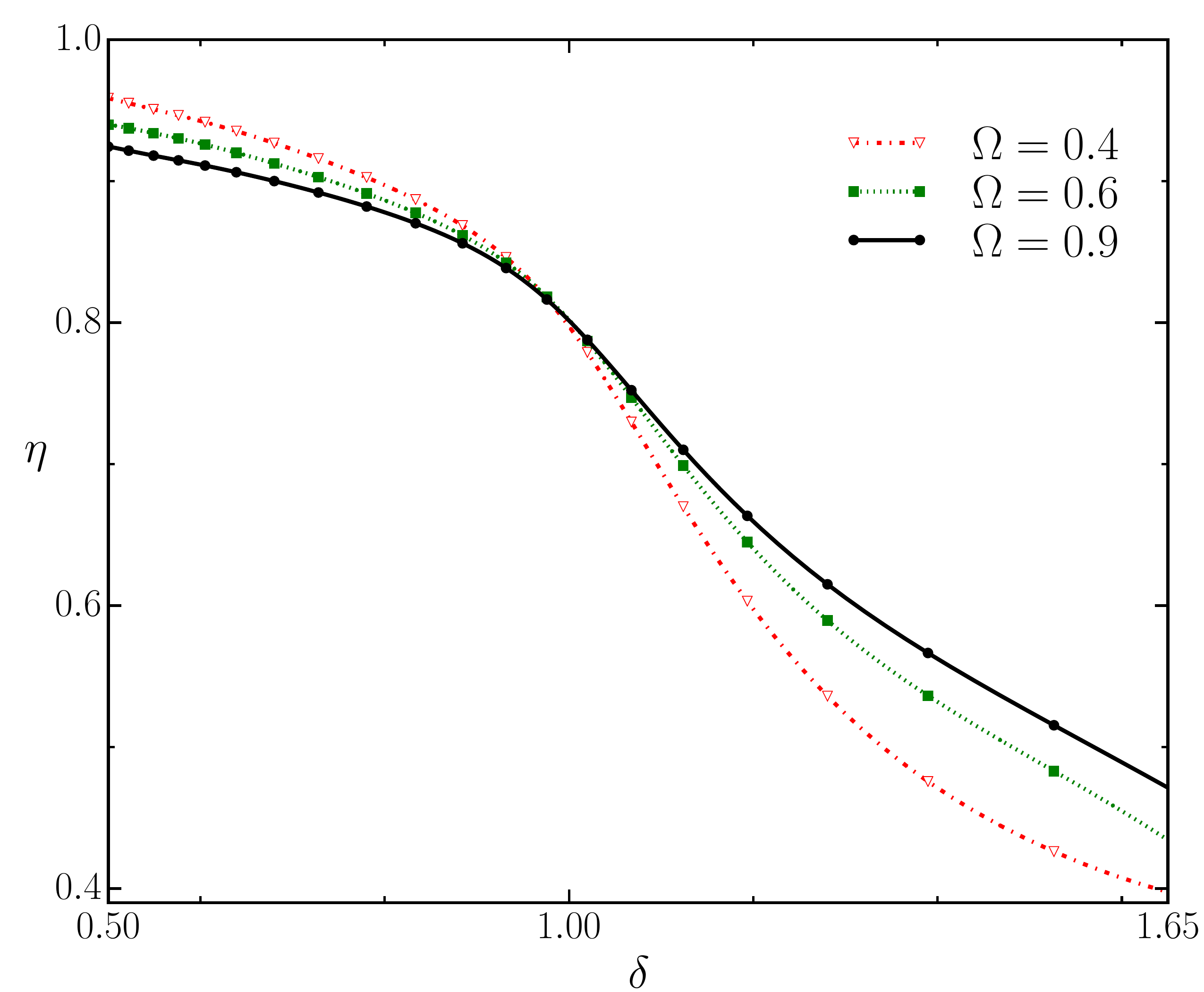}
\caption{Miscibility parameter $\protect\eta $ [Eq. (\protect\ref{eta})] is ploted
in terms of $\protect\delta$ for the $^{164}$Dy-$^{162}$Dy mixture at different
values of the rotation frequency $\Omega $.}
\label{fig7}
\end{figure}

In the well-miscible state, at $\delta <0.9$, triangular and square-shaped
vortex lattices are found as stable patterns. With the onset of
immiscibility, positions of vortices in one component shifted with respect
to the other, which leads to a transition in the respective lattice
structure. Namely, at $0.9<\delta <1.02$ the square-shaped lattice is
transformed into the rectangle one, as the vortices in each component tend
to get closer and form stripes. Therefore, in this regime, the system
produces rectangular and double-core vortices. Thus, in the immiscible state
at $\delta >1$, we observe vortex stripes, and also patterns that may be
called domain wall, see the panels (f$_j$) of Fig. \ref{fig4}.  
Actually, the phase diagram displayed in Fig. \ref{fig3} is similar to its counterpart
produced for non-dipolar BEC in Ref.~\cite{vortex-phase}, as well as to the
phase diagram for non-dipolar condensates produced~in Ref. \cite{Mueller} on
the basis of the lowest-Landau-level approximation. This similarity is
explained by the fact that, in the nearly-symmetric $^{164}$Dy-$^{162}$Dy
binary system, with equal dipole moments of both species, effects of equal
intra- and inter-species DDI on the miscibility almost cancel.

In Figs.~\ref{fig4} and \ref{fig5}, by considering the $^{164}$Dy-$^{162}$Dy mixture,
we display typical density plots for vortex lattices with different patterns, such as 
triangular, squared, rectangular, stripes and with domain walls, according to 
values of $\delta$, which can correspond to miscible or immiscible cases.
In Fig.~\ref{fig4}, with $\Omega =0.4$, one can observe that 
the stripe pattern forms overlapping lines of vortices in both component. 
In the double-core structure, vortex lattice in the second component is formed by pairs of
vortices with the same circulation, and vortices in the first component
surrounded by those pairs. In the strongly phase-separated regime, at $\delta >1$, 
vortices in one component are located very closely, merging into the
domain wall, with the walls in the two components being mutually interlaced.
In the Fig.~\ref{fig5} we consider the same parameters as in Fig.~\ref{fig4}, except that
the rotation frequency is changed to $\Omega =0.6$, in order to verify how 
the lattice shapes are being affected by $\Omega$.
In Fig.~\ref{fig6}, further examples for the $^{164}$Dy-$^{162}$Dy mixture
are displayed of striped vortices and domain walls, with  
$\delta >1$, by considering a large value $\Omega =0.8$ of the rotation
frequency.

The Thomas-Fermi (TF) density distribution for the overlapping binary BECs
subject to the solid rotation was verified in Ref.~\cite{vortex-phase}
to be a good approximation to the corresponding total density
distribution $\boldsymbol{n}_{T}=|\psi _{1}|^{2}+|\psi _{2}|^{2}$.
Due to the repulsion between the species, a vortex in one component
corresponds to a density peak in the other, and vice versa. 
In the present work, we have confirmed that the TF expression, given in 
Ref.~\cite{vortex-phase}, which can also be  derived by following the
lines of Ref.~\cite{Thomas},
$\boldsymbol{n}_{TF}(\rho)=2\sqrt{\alpha/\pi}-\alpha\rho^{2}$, 
holds also for the parameter regimes we are considering in the presence of 
DDI, with $\alpha\equiv {(1-\Omega ^{2})}/{(g_{11}+\mathrm{d}_{11}+g_{12}+\mathrm{d}_{12})}$. 
The agreement was confirmed for different miscible dipolar BEC mixtures.

In Fig.~\ref{fig7}, the miscible structure of the binary system is
illustrated by the dependence of parameter $\eta$, defined by Eq. (\ref{eta}), 
on the ratio between inter- and intra scattering-lengths as defined by 
Eq. (\ref{delta}), for different values of $\Omega $. 

\begin{figure}[tbp]
\centering\includegraphics[width=0.8\columnwidth]{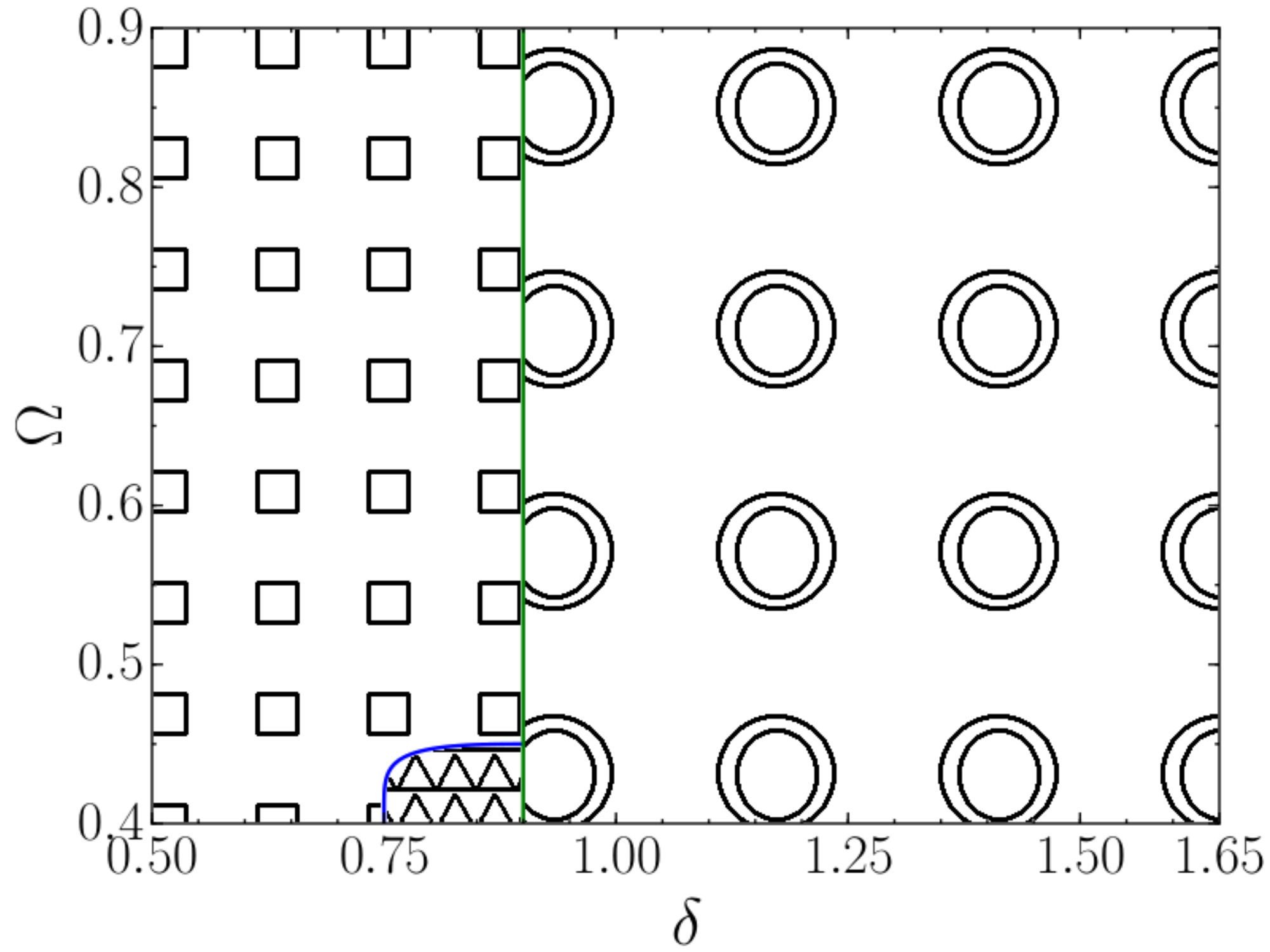}
\caption{The phase diagram of vortex patterns in the plane defined
by $\delta $ and the rotation frequency $\Omega$ plane, for the asymmetric 
$^{168}$Er-$^{164}$Dy mixture. The symbols for the observed lattice patterns are:
triangles for triangular-shaped, squares for squared-shaped, concentric
circles for circular lattices.}
\label{fig8}
\end{figure}

\begin{figure}[tbp]
\centering\includegraphics[width=0.5\textwidth]{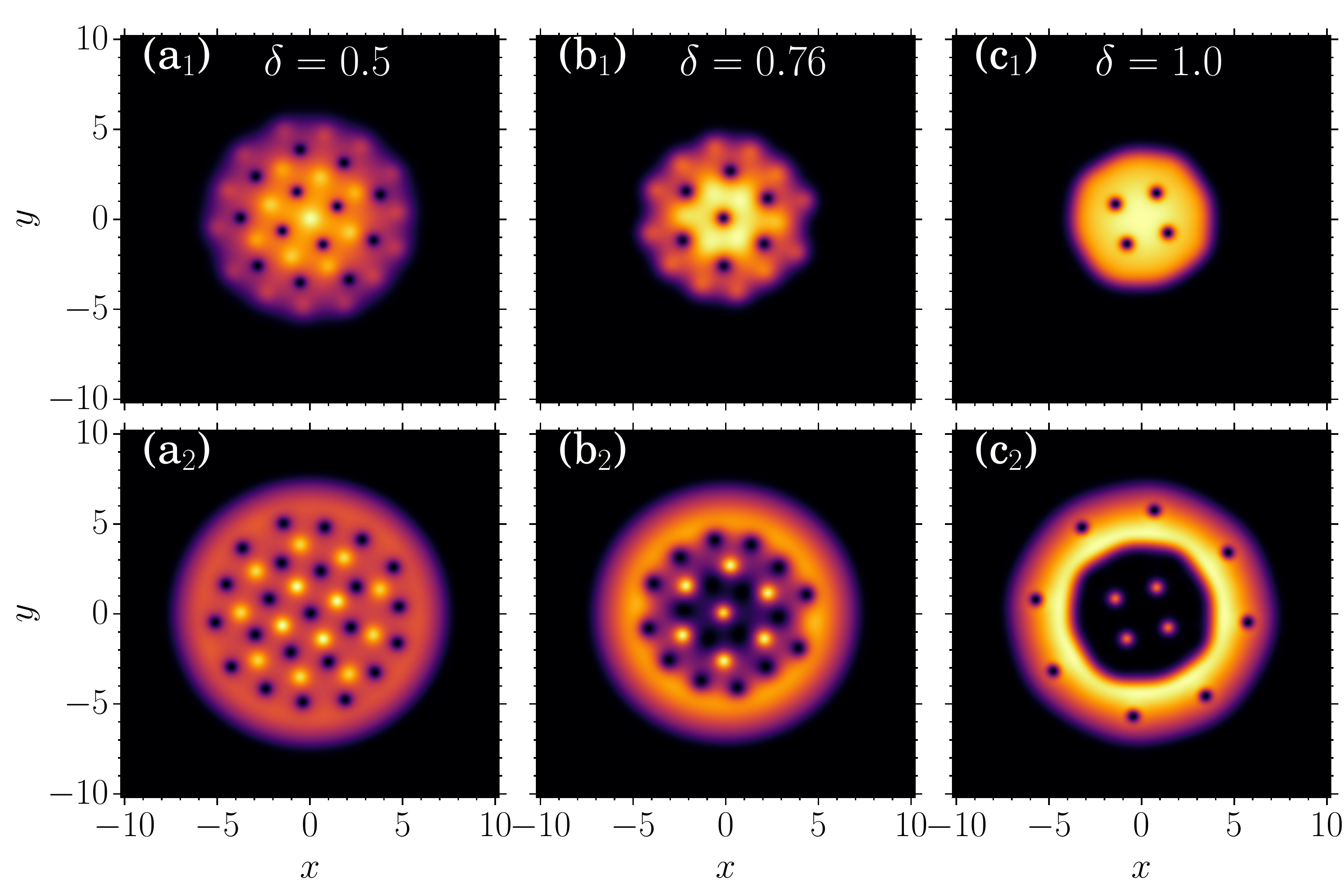}
\caption{Similar as in Fig.~\ref{fig4}, for the $^{168} $Er-$^{164}$Dy mixture, 
we have the 2D densities $|\protect\psi _{j}|^{2}$, for several values of $\protect\delta $.  
As in Fig.~\ref{fig4}, we have the rotation frequency $\Omega =0.4$, with
the other parameters given by $N_{j=1,2}=10^{4}$, $\lambda =20$
and $a_{12}=50a_{0}$.}
\label{fig9}
\end{figure}

\begin{figure}[tbp]
\centering\includegraphics[width=0.5\textwidth]{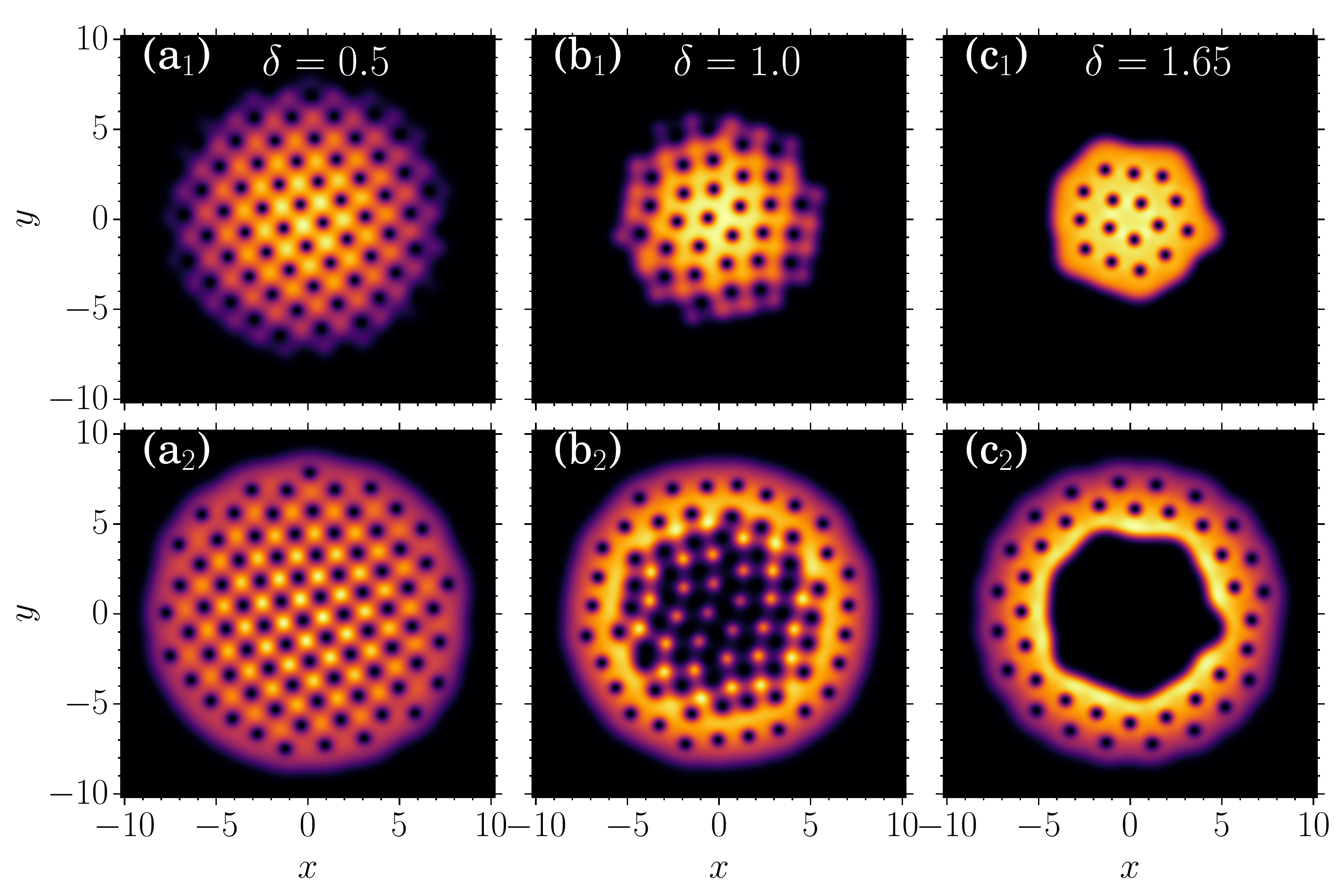}
\caption{The same as in Fig. \protect\ref{fig9}, with the rotation frequency changed to
$\Omega =0.8$.}
\label{fig10}
\end{figure}

\subsection{The asymmetric $^{168}$Er-$^{164}$Dy mixture under the action
of contact interactions}

The phase diagram of vortex patterns for parameters corresponding to the 
$^{168}$Er-$^{164}$Dy mixture is displayed in Fig.~\ref{fig8}. Recall that,
in the absence of contact interactions, this system is immiscible, as
shown above in Fig.~\ref{fig2}. However, imbalanced inter- and intra-species
contact interactions can impose miscibility in this setting. The phase
diagram of the asymmetric system is drastically different from that for its
symmetric $^{164}$Dy-$^{162}$Dy counterpart, due to the miscibility of the
latter in the absence of contact interactions, cf. Fig. \ref{fig3}. Note
that, while the immiscibility in non-dipolar binary condensates takes place
at $\delta >1$, in the $^{168}$Er-$^{164}$Dy system complete immiscibility
commences from $\delta =0.9$. The shift to $\delta <1$ is induced by the
imbalance of the DDI.

In case of immiscible states for the $^{168}$Er-$^{164}$Dy system at $\delta >0.9$, 
only circular-shaped lattice are established. 
In both the miscible and immiscible states, the squared- and circular-shaped  
lattices are shown in Figs.~\ref{fig9} and \ref{fig10}. Due to immiscibility, the first component
is surrounded by the second component, cf. Fig. \ref{fig2} which displays a
similar arrangement in the immiscible system of the $^{168}$Er-$^{164}$Dy
type. In such a phase-segregated mixture, vortices in the second component
are arranged into a circular lattice. A similar situation may also
occur in asymmetric non-dipolar systems with $a_{11}\neq a_{22}$ and $a_{12}=
\sqrt{a_{11}a_{22}}$, when unequal intra-species interactions may balance
the inter-species repulsion~\cite{2c-vort}.

\begin{figure}[tbp]
\centering\includegraphics[width=0.4\textwidth]{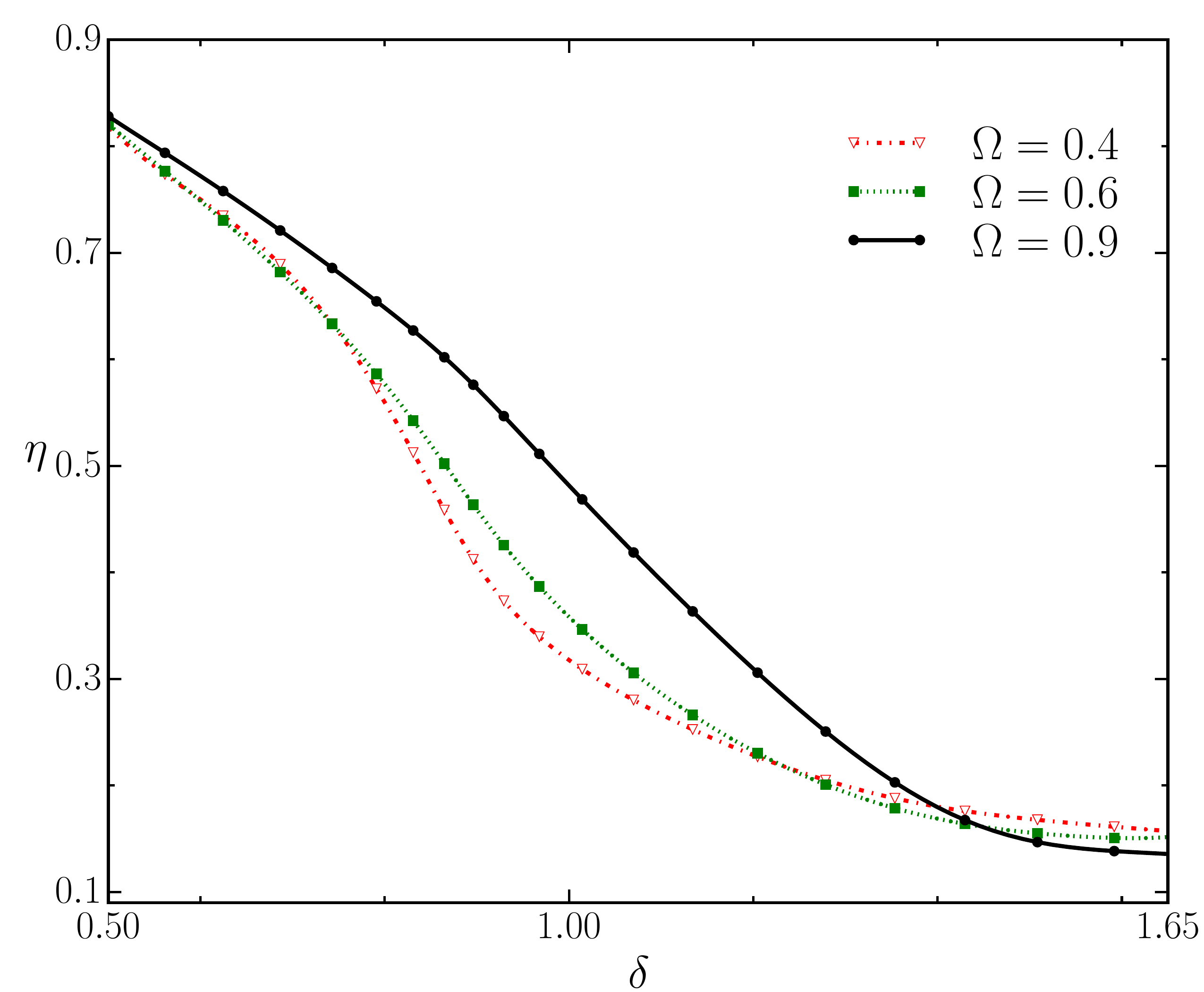}
\caption{The miscibility parameter $\protect\eta $ [Eq. (\protect\ref
{eta} )] as a function of $\protect\delta=a_{12}/a_{11}$ (with $a_{11}=a_{22}$ and $a_{12}=50a_0$) 
for the $^{168}$Er-$^{164}$Dy mixture, 
is represented for three different values of the rotation frequency, $\Omega $, 
as indicated.}
\label{fig11}
\end{figure}

In Fig.~\ref{fig11}, where the miscibility parameter $\eta $ is shown as a function of the 
scattering-length ratio $\delta $ [see Eq. (\ref{delta})], one can verify that the behavior obtained 
for the asymmetric $^{168}$Er-$^{164}$Dy mixture is quite different from the nearly-symmetric 
$^{164}$Dy-$^{162}$Dy mixture shown in Fig. \ref{fig7}.  By considering the same scattering lengths 
for both species ($\delta\approx 1$) we have the large miscibility of the dipolar $^{164}$Dy-$^{162}$Dy 
mixture (with $\eta\approx 0.8$) not too much affected by the rotation. However, in the same condition 
($\delta\approx 1$), the miscibility of the dipolar asymmetric $^{168}$Er-$^{164}$Dy mixture is strongly 
affected by the rotation: faster rotation increases the miscibility of the mixture.
As a general trend, for a large range of values for $\delta$, faster rotations tend to enhance 
mixing (increasing $\eta$) of the asymmetric dipolar mixture. This behavior is reversed only 
for $\delta\gtrapprox 1.5$. When the inter-species scattering length $a_{12}$ is about 1.5$a_{11}$ 
or larger, the miscibility decreases for larger rotations. In Table I, we can better verify the dependence of 
the miscibility on changes of the rotation parameter, for different values of $\delta$, considering the
two mixtures we are studying, as well as a case where one of the species is non-dipolar,
which is discussed in the following sub-section.

\begin{table}[th]
\caption{Miscibility parameter $\protect\eta $ [Eq. (\protect\ref{eta})] as a function of the inter-intra scattering-length 
ratio, $\protect\delta\equiv a_{12}/a_{11}$, at different values of the
rotation frequency, $\Omega $, for the mixtures considered in the
paper. For the sake of comparison in our analysis, a mixture with dipolar and non-dipolar species, $^{164}$Dy-$^{85}$Rb, 
is also included.}
\label{table1}
\begin{center}
\begin{tabular*}{1.0\columnwidth}{@{\extracolsep{\fill}}p{1.5cm}p{1.5cm}p{2cm}p{2cm}p{2cm}}
\hline\hline
\multicolumn{1}{p{1.5cm}}{$\delta$} & \multicolumn{1}{p{1.5cm}}{$
\Omega$} & \multicolumn{1}{p{2.2cm}}{$\eta$} & \multicolumn{1}{p{2.2cm}}{$
\eta$} & \multicolumn{1}{p{2.2cm}}{$\eta$} \\[1.5mm]
&  & \hspace{-0.95cm}$^{164}$Dy-$^{162}$Dy & \hspace{-0.95cm}$^{168}$Er-$
^{164}$Dy & \hspace{-0.95cm}$^{164}$Dy-$^{85}$Rb \\[2.5mm] \hline
0.5 & 0.0 & 1.0 & 0.86 & 0.93 \\
& 0.4 & 0.95 & 0.81 & 0.83 \\
& 0.9 & 0.92 & 0.82 & 0.82 \\[2.5mm] \hline
1.0 & 0.0 & 0.99 & 0.32 & 0.73 \\
& 0.4 & 0.83 & 0.30 & 0.69 \\
& 0.9 & 0.83 & 0.48 & 0.69 \\[2.5mm] \hline
1.5 & 0.0 & 0.23 & 0.17 & 0.59 \\
& 0.4 & 0.39 & 0.15 & 0.53 \\
& 0.9 & 0.46 & 0.13 & 0.54 \\[1.5mm] \hline\hline
\end{tabular*}
\end{center}
\end{table}

\subsection{The mixture of dipolar and non-dipolar species}

In this sub-section we comment briefly the coupled $^{164}$Dy-$^{85}$Rb mixture, where the magnetic moment 
of the second species ($^{85}$Rb) is negligible. This type of mixtures was discussed in Ref.~\cite{half-quantum}, 
where half-quantum vortex chains were reported, such that here we just include our corresponding results
in Table \ref{table1}, for the sake of comparison with the results we have obtained for the symmetric and 
asymmetric dipolar mixtures. We summarize in Table \ref{table1} our results for the dependence of the miscibility 
parameter $\eta $ on the relevant control parameters, which have being considered in the present work; i.e., 
the scattering-length ratio $\delta $ [Eq. (\ref{delta})] and rotation frequency $\Omega $. 

As a general remark, it is observed from the two components mixture of the $^{164}$Dy-$^{85}$Rb type,
that the absence of the DD repulsion between the two components tends to make the mixture less 
miscible in the presence of the rotation. 
When there is no-rotation, it is clear the strong role of the ratio between the scattering lengths, 
with the miscibility decreasing significantly for larger values of $\delta$. This trend is attenuated by
increasing the rotation of the coupled systems, as faster rotations tend to enhance mixing.
Another remark is that always the asymmetric dipolar mixture, $^{168}$Er-$^{164}$Dy,
is less miscible than the others two mixtures. 

\section{Summary and Conclusion}
\label{secIV}
Stimulated by the current interest from experiments with dipolar BEC mixtures, we have developed the 
detailed theoretical analysis for two co-rotating mixtures of this type; one, nearly symmetric (corresponding 
to parameters of the $^{164}$Dy-$^{162}$Dy system), and one asymmetric, which
represents the $^{168}$Er-$^{164}$Dy mixture. The dynamics of the mixtures
is described by the effectively 2D system of coupled GPEs (Gross-Pitaevskii
equations), which was derived from the full 3D system under the assumption
of strong confinement in the transverse direction. For that we consider pancake-type
condensates with aspect ratio given by $\lambda=$20.
The coupled equations include both the repulsive dipole-dipole interactions and repulsive contact
interactions, with our results being presented in two-dimensional density plots, complemented
by phase diagrams analysis related the two main parameters found for the miscibility of the species:
the rotation angular parameter $\Omega $ and the ratio between scattering lengths for the inter-
and intra-species contact interactions, given by $\delta$.
The phase diagrams display stability regions for several basic types of binary vortex lattices. 
In the absence of the contact interactions, the symmetric system is miscible, while the 
asymmetric one is not. The addition of contact interactions can change significantly the
situation. For the symmetric mixture, $^{164}$Dy-$^{162}$Dy, the phase diagram is similar to 
those recently found for binary non-dipolar condensates. It includes regions supporting vortex 
lattices with triangular, square-shaped, rectangular-shaped, double core, striped, and with domain
walls. The phase diagram for the asymmetric mixture, $^{168}$Er-$^{164}$Dy,
includes triangular, square-shaped, and circular lattices.

To understand the origin of the observed vortex patterns, it is relevant to recall a previous work~\cite{2c-compet-dip}, 
where vortices in a dipolar-nondipolar mixture were considered. It was found that the role of the dipolar component is 
to create vortices when the long-range dipolar interactions dominate over the contact nonlinearity. Following the pattern, 
we start the presentation of our results by considering pure dipolar mixtures, with both components involved
in the dipolar interactions. In this case, the long-range interactions give rise to two distinct kinds of vortex patterns, 
displayed in Figs.~\ref{fig1} and \ref{fig2}, the selection of a particular one being mainly determined by miscibility or 
immiscibility of the two-component system. The miscible system favors the square-shaped or striped lattices, whereas
the immiscibility tends to establish a hexagonal lattice in one component, surrounded by a ring-shaped structure in the other. 
It is concluded from the consideration of the settings which include contact interactions that,
in addition to the rotation frequency, the shape of the observed patterns
is strongly affected by the (im)miscibility of the coupled system, which may be
effectively shifted by the contribution from the contact interactions.

By summarizing the net effect of rotation, as well as contact interactions, in the miscibility of 
dipolar coupled systems, complementing the analysis presented in the Figs.~\ref{fig7} and 
\ref{fig11}, we include the Table \ref{table1}, for three values of the rotation parameter
$\Omega$ and three values of the scattering-ratio parameter $\delta$. In this table, for the sake of 
comparison with the dipolar systems we have studied, we also add results obtained for a non-dipolar 
coupled system, the $^{164}$Dy-$^{85}$Rb mixture, where the magnetic moment of the second 
species ($^{85}$Rb) is negligible. 
As noticed, by increasing the rotation, the coupled system becomes less miscible.
The strong role of the ratio between the scattering lengths $\delta$ for the miscibility can be clearly
verified from the results given in the table, with the parameter $\eta$ decreasing significantly 
as we increase this ratio. The general trend of the rotation is to attenuate such effect by
increasing the rotation of the coupled systems. 

To summarize, we have presented results on vortex-lattice structures expected to be of general interest 
in studies with dipolar mixtures. By considering particular dipolar mixtures, in specific pancake-type geometry,
we are contemplating dipolar BEC systems  in stable configurations, which are under active investigations
in cold-atom laboratories, with promising potential realization. 
Possible extensions of the present work on rotating binary condensates could be by including 
spin-orbit coupling effects, following analysis also studied in Refs.~\cite{Sakaguchi}. 
Another challenging generalization can be by studying spatially anisotropic quasi-2D configurations, with the 
magnetic dipoles oriented not perpendicularly to  the system's plane, but rather in the surface, considering 
that bright solitons were verified in such a configuration~\cite{Igor}. 

\acknowledgments RKK acknowledges the financial support from FAPESP of
Brazil (Contract number 2014/01668-8). AG and LT thank CAPES, CNPq and
FAPESP of Brazil for partial support. LT is also partially supported by
INCT-FNA (Proc. No. 464898/2014-5).
The work of BAM is partly supported
by a grant No. 2015616 from the joint program of the National Science
Foundation (US) and Binational Science Foundation (US-Israel), and by a
grant No. 1287/17 from the Israel Science Foundation.

\end{document}